\documentclass[12pt]{article}

\usepackage{amsmath}
\usepackage{amssymb}
\usepackage{slashed}
\usepackage[numbers,sort&compress]{natbib}
\usepackage{hyperref}
\usepackage{cleveref}
\crefname{equation}{Eq.}{Eqs.}
\crefname{figure}{Fig.}{Figs.}
\crefname{table}{Table}{Tables}
\crefname{section}{Section}{Sections}

\newcommand\scalemath[2]{\scalebox{#1} {\mbox{\ensuremath{\displaystyle #2}}}}
\usepackage[letterpaper,margin=1in,bottom=1in]{geometry}
\usepackage{float} 
\usepackage{parskip} 
\usepackage{tabulary} 
\usepackage{color} 
\usepackage{soul} 
\usepackage{subfigure}
\usepackage{graphicx}
\usepackage[section]{placeins} 

\def\lhc2{LHC~Run~II}

\usepackage{cleveref}

\newcommand{\code}[1]{\texttt{#1}}

\def\.4{\vspace{-.5cm}}
\newcommand{\ifb}{~\textrm{fb}^{-1}}

\def\beq{\begin{equation}}
\def\be{\begin{equation}}
\def\beqn{\begin{eqnarray}}
\def\ee{\end{equation}}
\def\eeq{\end{equation}}
\def\eeqn{\end{eqnarray}}

\author{
Amin Aboubrahim\footnote{Email: a.abouibrahim@northeastern.edu}~\ and 
Pran Nath\footnote{Email: p.nath@northeastern.edu}\\~\\
Department of Physics, Northeastern University,
Boston, MA 02115-5000, USA
}

\title{Detecting hidden sector dark matter at HL-LHC and HE-LHC via long-lived stau decays}
\date{}

\begin{document}
\maketitle

\vspace{1cm}

\textbf{Abstract: } 
 
We investigate a class of models where the supergravity model with the standard model gauge group is extended by a hidden sector $U(1)_X$ gauge group and where the lightest supersymmetric particle is the neutralino in the hidden sector. We investigate this possibility in a class of models where the stau is the lightest supersymmetric particle in the MSSM sector and the next-to-lightest supersymmetric particle of the $U(1)_X$-extended SUGRA model. In this case the stau will decay into the neutralino of the hidden sector. For the case when the mass gap between the stau and the hidden sector  neutralino is small and the mixing between the $U(1)_Y$ and $U(1)_X$ is also small, the stau can decay into the hidden sector neutralino and a tau which may be reconstructed as a displaced track coming from a high $p_T$ track of the charged stau. Simulations for this possibility are carried out for HL-LHC and HE-LHC. The discovery of such a displaced track from a stau will indicate the presence of hidden sector dark matter.

\newpage

\section{Introduction}\label{sec:intro}

Most of the searches for dark matter (DM) are focused on dark matter being a particle interacting weakly with the standard model (SM) particles and having a cross section in a range accessible to direct detection and indirect detection experiment.  For example in the context of supersymmetry (SUSY) if the lightest particle is neutral with R parity conservation, it is a candidate for dark matter. However, it is entirely possible that dark matter resides in hidden sectors which are ubiquitous in supergravity (SUGRA) and string models (see, e.g.,~\cite{Nath:2016qzm}). Further, SUGRA models with a minimal supersymmetric standard model (MSSM) spectrum extended by a $U(1)_X $ gauge group brings in an additional vector superfield with particle content of 
$B_\mu', \lambda_{X}$ where $B_\mu'$ is the new gauge boson and $\lambda_{X}$ is its gaugino superpartner.
 The $U(1)_X$ can mix with hypercharge  $U(1)_Y$ via kinetic mixing~\cite{Holdom:1985ag, Holdom:1991}. 
 Additionally with Stueckelberg mass mixing of 
 $U(1)_X$ and $U(1)_Y$ one brings in a chiral superfield which contains a Weyl fermion $\psi$~\cite{Kors:Nath,st-mass-mixing,Feldman:2007wj}.
  After electroweak symmetry 
 breaking the above leads to a $6\times 6$ neutralino mass matrix, where the additional two neutralinos reside in the hidden
 sector with highly suppressed couplings to the visible sector. Let us suppose that  one of the two neutralinos which lie
 in the hidden sector is the lightest supersymmetric particle (LSP) of the extended model and further the next-to-lightest supersymmetric particle (NLSP) is a stau which lies close to the hidden sector
 neutralino. In this case the stau will decay into the hidden sector neutralino with a long lifetime. Such a decay can leave a
 track in the inner detectors (ID) of the ATLAS and CMS experiments. In this work we explore this possibility within the 
 framework of supergravity  grand unified model
 with an extended $U(1)_X$ sector including both the gauge kinetic mixing and
 the Stueckelberg mass mixing. $U(1)$ extensions of supersymmetric models and their implications on dark matter and collider analyses have been studied extensively in the literature~\cite{U1extensions}. However, the setup in the present work  is quite different from these.
 
 The outline of the rest of paper is as follows: In section~\ref{sec:model} we discuss the $U(1)_X$ extended SUGRA model
 with  gauge kinetic mixing and Stueckelberg mass mixing. 
 In section~\ref{sec:implement}, we discuss implementation of this model and the mechanism that leads to a long lived stau consistent with the current experimental constraints on
 the light Higgs boson mass as measured by the ATLAS and CMS     Collaborations~\cite{Aad:2012tfa,Chatrchyan:2012ufa},
 and the relic density as measured by the Planck Collaboration~\cite{Aghanim:2018eyx}. 
 In section~\ref{sec:DM}, further details of the generation of relic density for the 
 dark matter in the hidden sector is discussed. 
 Currently, the LHC has completed its phase 2 and has shut down for two years for  the period 2019-2020 for 
 an upgrade and the upgraded LHC  will operate at 14 TeV in the period 2021-2023.  During this period the upgraded LHC 
 will collect about  300 fb$^{-1}$ of additional data for each detector.
 Thereafter there will be a major upgrade of the LHC to 
 high luminosity LHC (HL-LHC) during the period 2023-2026. This final upgraded HL-LHC will resume operations in late 2026
 and is expected to run for ten years till 2036. It is projected that at the end of this period  each detector will collect
  about 3000 fb$^{-1}$ of data.
  Future colliders beyond HL-LHC are also  being  discussed.  Among these are 
 a 100 TeV  $pp$  collider  at CERN 
 and  also a  100 TeV  $pp$  collider in China~\cite{Arkani-Hamed:2015vfh,Mangano:2017tke}
  each of which requires a circular ring of about 100 km.   
  Further, a third possibility of a 27 TeV $pp$ collider, the high energy LHC (HE-LHC) at 
 CERN is also under study~\cite{CidVidal:2018eel,Cepeda:2019klc, Benedikt:2018ofy,Zimmermann:2018koi}.
 Such a collider can be built  within  the existing tunnel at CERN
by installing 16 T 
superconducting magnets  using FCC technology
capable of enhancing the center-of-mass energy of the collider to 27 TeV. 
If built, the HE-LHC  will operate at a luminosity of $2.5\times 10^{35}$ cm$^{-2}$s$^{-1}$ 
and collect 10$-$15 ab$^{-1}$ of data. In this work we will focus on HL-LHC and HE-LHC.
 Thus in section~\ref{sec:production}, we discuss the production cross section of the  NLSP stau 
 at the LHC at 14 TeV and at 27 TeV (for previous work on HL-LHC and HE-LHC see~\cite{Aboubrahim:2018tpf,Aboubrahim:2018bil,Aboubrahim:2017wjl,Aboubrahim:2017aen}).
 In section~\ref{sec:simulation}, an analysis of signal and background simulation and event
 selection is carried out.  In section~\ref{sec:results}, a cut-flow analysis and and the result of this analysis  are discussed. Here the analysis is done
 with no pile-up and with pile-up. The analysis also makes a comparative study of the discovery potential of HL-LHC 
 and HE-LHC for the detection of hidden sector dark matter. Conclusions are given in section~\ref{sec:conc}.

\section{The model}\label{sec:model}

As discussed above  we consider an extension  of the standard model gauge group by 
an additional abelian gauge group $U(1)_X$ of gauge coupling strength $g_X$. The particle spectrum  in the visible sector, i.e., quarks, leptons, Higgs and their superpartners are assumed neutral under $U(1)_X$. We focus first on the abelian gauge sector of the extended model which contains two vector superfields, a vector superfield $B$ associated with the hypercharge gauge group $U(1)_Y$, a vector superfield $C$ associated with the hidden sector gauge group $U(1)_X$, and a chiral scalar superfield $S$. In the Wess-Zumino gauge the $B$ and $C$ superfields have the following components 
\begin{equation}
B=-\theta\sigma^{\mu}\bar{\theta}B_{\mu}+i\theta\theta\bar{\theta}\bar{\lambda}_B-i\bar{\theta}\bar{\theta}\theta\lambda_B+\frac{1}{2}\theta\theta\bar{\theta}\bar{\theta}D_B,
\end{equation}
and
\begin{align}
C=-\theta\sigma^{\mu}\bar{\theta}C_{\mu}+i\theta\theta\bar{\theta}\bar{\lambda}_{C}-i\bar{\theta}\bar{\theta}\theta\lambda_{C}+\frac{1}{2}\theta\theta\bar{\theta}\bar{\theta}D_{C}.
\end{align}
The chiral scalar superfield  $S$ has the expansion
\begin{align}
S = &\frac{1}{2}(\rho+i a)+\theta\chi+i\theta\sigma^{\mu}\bar{\theta}\frac{1}{2}(\partial_{\mu}\rho+i\partial_{\mu}a) \\ \nonumber
&+\theta\theta F+\frac{i}{2}\theta\theta\bar{\theta}\bar{\sigma}^{\mu}\partial_{\mu}\chi+\frac{1}{8}\theta\theta\bar{\theta}\bar{\theta}(\square\rho+i\square a).
\end{align}
  
The gauge kinetic energy  sector of the model is
\begin{equation}
\mathcal{L}_{\rm gk}=-\frac{1}{4}(B_{\mu\nu}B^{\mu\nu}+C_{\mu\nu}C^{\mu\nu})-i\lambda_B\sigma^{\mu}\partial_{\mu}\bar{\lambda}_B-i\lambda_{C}\sigma^{\mu}\partial_{\mu}\bar{\lambda}_{C}+\frac{1}{2}(D^2_B+D^2_{C}).
\label{kinetic-1}
\end{equation} 
Next we allow  gauge kinetic mixing between the $U(1)_X$ and $U(1)_Y$ sectors with terms of the form
\begin{equation}
-\frac{\delta}{2}B^{\mu\nu}C_{\mu\nu}-i\delta(\lambda_{C}\sigma^{\mu}\partial_{\mu}\bar{\lambda}_B+\lambda_{B}\sigma^{\mu}\partial_{\mu}\bar{\lambda}_{C})+\delta D_B D_{C}.
\label{kinetic-2}
\end{equation}

As a result of Eq.~(\ref{kinetic-1}) the hidden $U(1)_X$ interacts with the MSSM fields via the small kinetic mixing parameter $\delta$. The kinetic terms in Eq.~(\ref{kinetic-1}) and Eq.~(\ref{kinetic-2})  can be diagonalized using the transformation
\beqn
\left(\begin{matrix} B^{\mu} \cr 
C^{\mu} 
\end{matrix}\right) = \left(\begin{matrix} 1 & -s_{\delta} \cr 
0 & c_{\delta} 
\end{matrix}\right)\left(\begin{matrix} B'^{\mu} \cr 
C'^{\mu} 
\end{matrix}\right), 
\label{rotation}
\eeqn  
where $c_{\delta}=1/(1-\delta^2)^{1/2}$ and $s_{\delta}=\delta/(1-\delta^2)^{1/2}$. 

Aside from gauge kinetic mixing, we assume a Stueckelberg mass mixing between the $U(1)_X$ and $U(1)_Y$ sectors so that 

\begin{equation}
\mathcal{L}_{\rm St}=\int d\theta^2 d\bar{\theta}^2(M_1 C+M_2 B+S+\bar{S})^2.
\label{lag}
\end{equation}  
We note that Eq.~(\ref{lag}) is invariant under $U(1)_Y$ and $U(1)_X$ gauge transformation so that,
\begin{align}
&\delta_Y B = \Lambda_Y+\bar{\Lambda}_Y, \, \, \, \, \, \delta_Y S = -M_2\Lambda_Y, \\ \nonumber 
& \delta_X C = \Lambda_X+\bar{\Lambda}_X, \, \, \, \delta_X S = -M_1\Lambda_X.
\end{align}
In component notation, $\mathcal{L}_{\rm St}$ is 
\begin{align}
\mathcal{L}_{\rm St} = &-\frac{1}{2}(M_1 C_{\mu}+M_2 B_{\mu}+\partial_{\mu}a)^2-\frac{1}{2}(\partial_{\mu}\rho)^2-i\chi\sigma^{\mu}\partial_{\mu}\bar{\chi}+2|F|^2 \\ \nonumber
&+\rho(M_1 D_{C}+M_2 D_B)+\bar{\chi}(M_1\bar{\lambda}_{C}+M_2\bar{\lambda}_B)+\chi(M_1\lambda_{C}+M_2\lambda_B). 
\end{align}
In unitary gauge the axion field $a$ is absorbed to generate mass for the $U(1)_X$ gauge boson.  \\
It is convenient from this point on to introduce Majorana spinors $\psi_S$, $\lambda_X$ and $\lambda_Y$ so that   
 \begin{equation}
  \psi_S =
  \begin{pmatrix}
    \chi_{\alpha}  \\
    \bar{\chi}^{\dot{\alpha}} 
  \end{pmatrix},\quad
  \lambda_X=
  \begin{pmatrix}
    \lambda_{C\alpha}  \\
    \bar{\lambda}^{\dot{\alpha}}_{C} 
  \end{pmatrix},\quad
  \lambda_Y=
  \begin{pmatrix}
    \lambda_{B\alpha}  \\
    \bar{\lambda}^{\dot{\alpha}}_{B}
  \end{pmatrix}.
  \label{spinors}
\end{equation}
In addition to the above we add a soft SUSY breaking term to the Lagrangian so that 
\begin{equation}
\Delta\mathcal{L}_{\rm soft} \
=-\left(\frac{1}{2}m_X\bar{\lambda}_X\lambda_X+ M_{XY}\bar{\lambda}_X\lambda_Y\right)-\frac{1}{2} m^2_{\rho}\rho^2,
\end{equation}
where $m_X$ is mass of the $U(1)_X$ gaugino and $M_{XY}$ is the $U(1)_X$-$U(1)_Y$ mixing mass. 
We note that the mixing parameter $M_{XY}$ and $M_2$ even when set to zero at the grand unification scale will assume 
non-vanishing values due  to renormalization group evolution. Thus $M_{XY}$ has the beta-function evolution so that 
\begin{equation}
\beta^{(1)}_{M_{XY}}=\frac{33}{5}g^2_Y\left[M_{XY}-(M_1+m_X)s_{\delta}+M_{XY}s^2_{\delta} \right],
\end{equation}
where $g_Y$ is the $U(1)_Y$ gauge coupling. Similarly, the  mixing parameter $M_2$  
has the beta-function so that 
\begin{equation}
\beta^{(1)}_{M_2}=\frac{33}{5}g^2_Y(M_2-M_1 s_{\delta}).
\label{m2rge}
\end{equation}
In the MSSM sector we will take the soft terms to consist of $m_0, ~A_0, ~m_1, ~m_2, ~m_3, ~\tan\beta,
 ~\text{sgn} (\mu)$. Here 
 $m_0$ is the universal scalar mass, $A_0$ is the universal trilinear coupling, $m_1,  ~m_2,  ~m_3$ are the masses of the $U(1)$, $SU(2)_L$, and $SU(3)_C$ gauginos, $\tan\beta=v_u/v_d$ is the ratio of the Higgs VeVs and $\text{sgn}(\mu)$ is the sign of the Higgs mixing parameter which is chosen to be positive. 

We focus first on the neutralino sector of the extended SUGRA model. We choose as basis $(\psi_S,\lambda_X,\lambda_Y,\lambda_3,\tilde h_1, \tilde h_2)$ where the first two fields arise from the extended sector and the last four, i.e., 
 $\lambda_Y, \lambda_3, \tilde h_1, \tilde h_2$ are the gaugino and higgsino fields of the MSSM sector. Using Eq.~(\ref{rotation}) we rotate into the new basis $(\psi_S,\lambda'_X,\lambda'_Y,\lambda_3,\tilde h_1, \tilde h_2)$ so that the $6\times 6$ neutralino mass matrix takes the form
\beqn
\scalemath{0.9}{
\left(\begin{matrix}  0 & M_1 c_{\delta}-M_2 s_{\delta} & M_2 & 0 & 0 & 0 \cr
M_1 c_{\delta}-M_2 s_{\delta} & m_X c^2_{\delta}+m_1 s^2_{\delta}-2M_{XY}c_{\delta}s_{\delta} & -m_1 s_{\delta}+M_{XY}c_{\delta} & 0 & s_{\delta}c_{\beta}s_W M_Z & -s_{\delta}s_{\beta}s_W M_Z \cr
M_2 & -m_1 s_{\delta}+M_{XY}c_{\delta} & m_1 & 0 & -c_{\beta}s_W M_Z & s_{\beta}s_W M_Z \cr
0 & 0 & 0 & m_2 & c_{\beta}c_W M_Z & -s_{\beta}c_W M_Z \cr
0 & s_{\delta}c_{\beta}s_W M_Z & -c_{\beta}s_W M_Z & c_{\beta}c_W M_Z & 0 & -\mu \cr
0 & -s_{\delta}s_{\beta}s_W M_Z & s_{\beta}s_W M_Z & -s_{\beta}c_W M_Z & -\mu & 0 \cr
\end{matrix}\right)},
\eeqn 
where 
$s_{\beta}\equiv\sin\beta$, $c_{\beta}\equiv\cos\beta$, $s_W\equiv\sin\theta_W$, $c_W\equiv\cos\theta_W$ with $M_Z$ being the $Z$ boson mass. 
 We label the mass eigenstates as 
 \begin{equation}
 \tilde\xi^0_1, ~\tilde\xi^0_2; ~\tilde \chi_1^0, ~\tilde \chi_2^0, ~\tilde \chi_3^0, ~\tilde \chi_4^0\,.
 \end{equation}
Here the first two neutralinos  $\tilde\xi^0_1$ and $\tilde\xi^0_2$ reside mostly in the hidden sector while the remaining four $\tilde \chi_i^0$ 
($i=1\cdots 4$) reside mostly in the MSSM sector. We assume $\tilde\xi^0_1$ to be the LSP. In the limit of small mixings 
between the hidden and the MSSM sector the masses of the hidden sector neutralinos are 
\begin{equation}
m_{\tilde\xi^0_1}=\sqrt{M_1^2+\frac{1}{4}\tilde m^2_X}-\frac{1}{2}\tilde m_X, \quad \text{and} \quad m_{\tilde\xi^0_2}=\sqrt{M_1^2+\frac{1}{4}\tilde m^2_X}+\frac{1}{2}\tilde m_X.
\end{equation}
For the case when $\tilde \xi_1^0$ is the least massive of all sparticles in the  $U(1)_X$ extended SUGRA model, dark matter will reside
in the hidden sector. Such a possibility has been foreseen in previous works (see, e.g.,~\cite{Feldman:2006wd,Feldman:2008xs,Feldman:2009wv}).

We turn now to the charge neutral gauge vector boson sector. Here the $2\times 2$ mass square matrix of the standard model 
is enlarged to become a $3\times 3$ mass square matrix in the $U(1)_X$-extended SUGRA model.
Thus  after spontaneous electroweak symmetry breaking and  the Stueckelberg mass growth the 
$3\times 3$ mass squared matrix of neutral vector bosons in the basis $(C'_{\mu}, B'_{\mu}, A^3_{\mu})$ is given by
\beqn
\mathcal{M}^2_V=\left(\begin{matrix}  M_1^2\kappa^2+\frac{1}{4}g^2_Y v^2 s^2_{\delta} & M_1 M_2\kappa-\frac{1}{4}g^2_Y v^2 s_{\delta} & \frac{1}{4}g_Y g_2 v^2 s_{\delta} \cr
M_1 M_2\kappa-\frac{1}{4}g^2_Y v^2 s_{\delta} & M_2^2+\frac{1}{4}g^2_Y v^2 & -\frac{1}{4}g_Y g_2 v^2 \cr
\frac{1}{4}g_Y g_2 v^2 s_{\delta} & -\frac{1}{4}g_Y g_2 v^2 & \frac{1}{4}g^2_2 v^2 \cr
\end{matrix}\right),
\label{zmassmatrix}
\eeqn
where $A^3_{\mu}$ is the third isospin component, $g_2$ is the $SU(2)_L$ gauge coupling, $\kappa=(c_{\delta}-\epsilon s_{\delta})$, $\epsilon=M_2/M_1$ and $v^2=v^2_u+v^2_d$. The mass-squared matrix of Eq.~(\ref{zmassmatrix}) has one zero eigenvalue which is the photon while the other two eigenvalues are
\begin{align}
&M^2_{\pm} = \frac{1}{2}\Bigg[M_1^2\kappa^2+M^2_2+\frac{1}{4}v^2[g_Y^2 c^2_{\delta}+g_2^2] \nonumber \\
&\pm \sqrt{\left(M_1^2\kappa^2+M^2_2+\frac{1}{4}v^2[g_Y^2 c^2_{\delta}+g_2^2]\right)^2-\Big[M_1^2 g_2^2v^2\kappa^2+M_1^2g^2_Yv^2 c^2_{\delta}+M_2^2g^2_2 v^2\Big]}~\Bigg],
\label{bosons}
\end{align} 
where $M_+$ is identified as the $Z'$ boson mass while $M_-$ as  the $Z$ boson. The diagonalization of the mass-squared matrix of Eq.~(\ref{zmassmatrix}) can be done via two orthogonal transformations where the first is given by~\cite{Feldman:2007wj}
\beqn
\mathcal{O}=\left(\begin{matrix} 1/c_{\delta} & -s_{\delta}/c_{\delta} & 0 \cr
s_{\delta}/c_{\delta} & 1/c_{\delta} & 0 \cr
0 & 0 & 1 \cr
\end{matrix}\right),
\label{omatrix}
\eeqn
which transforms the mass matrix to $\mathcal{M'}^2_V=\mathcal{O}^{T}\mathcal{M}^2_V\mathcal{O}$, 
\beqn
\mathcal{M'}^2_V=\left(\begin{matrix}  M_1^2 & M_1^2\alpha & 0 \cr
M_1^2\alpha & M_1^2\alpha^2+\frac{1}{4}g^2_Y v^2 c^2_{\delta} & -\frac{1}{4}g_Y g_2 v^2 c_{\delta} \cr
0 & -\frac{1}{4}g_Y g_2 v^2 c_{\delta} & \frac{1}{4}g^2_2 v^2 \cr
\end{matrix}\right),
\label{zpmassmatrix}
\eeqn  
where $\alpha=\epsilon c_{\delta}-s_{\delta}$.
The gauge eigenstates of $\mathcal{M'}^2_V$ can be rotated into the corresponding mass eigenstates $(Z',Z,\gamma)$ using the second transformation via the rotation matrix
\beqn
\mathcal{R}=\left(\begin{matrix} c'_W c_{\phi}-s_{\theta}s_{\phi}s'_W & s'_W c_{\phi}+s_{\theta}s_{\phi}c'_W & -c_{\theta}s_{\phi} \cr
c'_W s_{\phi}+s_{\theta}c_{\phi}s'_W & s'_W s_{\phi}-s_{\theta}c_{\phi}c'_W & c_{\theta}c_{\phi} \cr
-c_{\theta} s'_W & c_{\theta} c'_W & s_{\theta} \cr
\end{matrix}\right),
\label{rotmatrix}
\eeqn
with $c'_W(c_{\theta})(c_{\phi})\equiv \cos\theta'_W(\cos\theta)(\cos\phi)$ and $s'_W(s_{\theta})(s_{\phi})\equiv \sin\theta'_W(\sin\theta)(\sin\phi)$, where $\theta'_W$ represents the mixing angle between the new gauge sector and the standard model gauge bosons while the other angles are given by
\begin{equation}
\tan\phi=\alpha, ~~~ \tan\theta=\frac{g_Y}{g_2}c_{\delta}\cos\phi,
\end{equation}
such that $\mathcal{R}^T\mathcal{M'}^2_V\mathcal{R}=\text{diag}(M^2_{Z'},M^2_{Z},0)$.
The resulting mixing angle is thus given by
\begin{equation}
\tan2\theta'_W\simeq\frac{2\alpha M^2_Z\sin\theta}{M^2_{Z'}-M^2_Z+(M^2_{Z'}+M^2_Z-M^2_W)\alpha^2},
\end{equation}
with $M_W=g_2 v/2$, $M_{Z'}\equiv M_+$ and $M_{Z}\equiv M_-$. 

\section{Model implementation and long-lived stau}\label{sec:implement}
One of the by-products of models with a hidden sector coupling to the MSSM only via a small kinetic mixing is the presence of long-lived particles (LLP) with late decays into hidden sector particles. The signature of the production of such particles at hadron colliders is very unique especially if the LLP is charged and leaves a track in the detector which can be easily identified. In this study we will be looking for long-lived staus which have lifetimes long enough allowing them to decay inside the detector tracker.

The model described in section~\ref{sec:model} is implemented in the mathematica package \code{SARAH-4.14}~\cite{Staub:2013tta,Staub:2015kfa} which generates model files for \code{SPheno-4.0.3}~\cite{Porod:2003um,Porod:2011nf} which in turn produces the sparticle spectrum and \code{CalcHep/CompHep}~\cite{Pukhov:2004ca,Boos:1994xb} files used by \code{micrOMEGAs-5.0.4}~\cite{Belanger:2014vza} to determine the dark matter relic density and \code{UFO} files~\cite{Degrande:2011ua} which are input to \code{MadGraph5}~\cite{Alwall:2014hca}. 

The input parameters of the $U(1)_X$-extended MSSM/SUGRA~\cite{msugra} are of the usual non-universal SUGRA model with additional parameters as below (all at the GUT scale)
\begin{equation}
m_0, ~~A_0, ~~ m_1, ~~ m_2, ~~ m_3, ~~M_1, ~~m_X, ~~\delta, ~~\tan\beta, ~~\text{sgn}(\mu),
\label{sugra}
\end{equation}    
where $m_0, ~A_0, ~m_1, ~m_2, ~m_3, ~\tan\beta$ and $\text{sgn}(\mu)$ are the soft parameters in the MSSM sector as defined earlier.  
The parameters $M_2$ and $M_{XY}$ are set to zero at the GUT scale. 
The input parameters must be such as to satisfy a number of experimental constraints. These include the constraint that the computed Higgs boson mass must be consistent with the Higgs boson mass measurements by the ATLAS and the CMS collaborations.
Further, the relic density of dark matter given by the model must be  consistent with that measured by the Planck experiment, and sparticle spectrum of the model be consistent with the lower 
experimental limits on sparticle masses. The consistency of the computed Higgs boson mass with the experimental determination 
of $m_{h^0}\sim 125$ GeV requires the loop correction to the Higgs boson mass be large which in turn implies that the 
size of weak scale supersymmetry lie in the several TeV region. Typically this leads to the average squark masses also lying
in the TeV region. Such a situation is realized on the hyperbolic branch of radiative breaking of electroweak 
symmetry~\cite{Chan:1997bi,Chattopadhyay:2003xi,Akula:2011jx}
(for related works see~\cite{Feng:1999mn,Baer:2003wx,Baer:2018hpb,Feldman:2011ud,Ross:2017kjc}).
It turns out that there are at least two ways in which the squark masses may be large, i.e., either $m_0$ is large or $m_3$ is large
lying in the several TeV region while $m_0$ can be relatively small. In the latter case renormalization group running would generate 
squark masses lying in the several TeV region while the slepton masses would be relatively much lighter~\cite{Akula:2013ioa}.
 In this analysis we follow the
second possibility and choose $m_3$ in the several TeV region but $m_0$ relatively much smaller.

With this set of input parameters, we scan the $U(1)_X$-extended MSSM/SUGRA parameter space to obtain a set of benchmark points satisyfing the Higgs boson mass at $125\pm 2$ GeV and the dark matter relic density at $\Omega h^2 \leq 0.123$. The benchmark points are shown in Table~\ref{tab1}.          

\begin{table}[H]
\begin{center}
\begin{tabulary}{0.85\textwidth}{l|CCCCCCCCC}
\hline\hline\rule{0pt}{3ex}
Model & $m_0$ & $A_0$ & $m_1$ & $m_2$ & $m_3$ & $M_1$ & $m_X$ & $\tan\beta$ & $\delta$ \\
\hline\rule{0pt}{3ex}  
\!\!(a)& 300 & 1838 & 885 & 740 & 4235 & 473 & 600 & 14 & $2.0\times10^{-5}$ \\
(b)  & 546 & -3733 & 828 & 761 & 3657 & 426 & 392 & 16 & $4.7\times10^{-6}$ \\
(c)  & 529 & -3211 & 864 & 482 & 3777 & 461 & 400 & 15 & $6.0\times10^{-6}$ \\
(d)  & 680 & -5198 & 1166 & 806 & 3945 & 503 & 198 & 15 & $2.5\times10^{-6}$ \\
(e)  & 563 & -1850 & 1214 & 598 & 3856 & 579 & 380 & 21 & $2.4\times10^{-6}$ \\
(f)  & 500 & -2698 & 1286 & 893 & 4165 & 523 & 65 & 15 & $2.5\times10^{-6}$ \\
(g)  & 515 & -261 & 1451 & 1265 & 4830 & 682 & 258 & 25 & $1.4\times10^{-6}$ \\
(h)  & 645 & 1009 & 1621 & 1160 & 5374 & 714 & 100 & 26 & $1.3\times10^{-6}$ \\
\hline
\end{tabulary}\end{center}
\caption{Input parameters for the benchmarks used in this analysis. Here $M_2=M_{XY}=0$ at the GUT scale. All masses are in GeV.}
\label{tab1}
\end{table}

We choose the parameters $m_1$, $m_2$, $M_1$ and $m_X$ so that the hidden sector neutralino $\tilde\xi^0_1$ is the LSP and thus the dark matter candidate. The small value for $m_0$ allows the stau to be the NLSP. However, the smallness of $m_0$ can be problematic for satisfying the Higgs boson mass. This is compensated by requiring a large $m_3$~\cite{Akula:2013ioa} as evident from Table~\ref{tab1}. The RGE running of the stop mass is driven by $m_3$ which develops a large enough mass to bring the Higgs mass above its tree-level value and close to the experimentally observed one. In the process, the gluino also gets a large mass. The resulting spectrum of some of the revelant particles is shown in Table~\ref{tab2}. 

\begin{table}[H]
\begin{center}
\begin{tabulary}{1.3\textwidth}{l|CCCCCCCCCCC}
\hline\hline\rule{0pt}{3ex}
Model  & $h^0$ & $\mu$ & $\tilde\chi_1^0$ & $\tilde\chi_1^\pm$ & $\tilde{\tau}$ & $\tilde{\nu}_{\tau}$ & $\tilde{\xi}^0_1$ & $\tilde t$ & $\tilde g$ & $\Omega h^2$ & $c\tau_0$ \\
\hline\rule{0pt}{3ex} 
\!\!(a) & 123.0 & 4127 & 359.9 & 556.9 & 275.1 & 434.3 & 260.1 & 6306 & 8459 & 0.116 & 243.6 \\
(b) & 123.1 & 4417 & 343.3 & 595.2 & 291.0 & 572.4 & 272.9 & 5118 & 7372 & 0.123 & 199.9 \\
(c) & 123.4 & 4426 & 350.3 & 350.5 & 319.3 & 459.8 & 302.5 & 5376 & 7621 & 0.109 & 147.0 \\
(d) & 124.6 & 4998 & 495.2 & 633.2 & 428.0 & 671.4 & 413.6 & 5347 & 7916 & 0.121 & 177.6 \\
(e) & 123.1 & 4236 & 449.0 & 449.2 & 440.5 & 570.6 & 419.4 & 5607 & 7764 & 0.111 & 307.6 \\
(f) & 124.2 & 4669 & 546.0 & 699.7 & 500.0 & 653.6 & 491.5 & 5926 & 8326 & 0.119 & 387.3 \\
(g) & 123.2 & 4852 & 619.4 & 1009 & 583.0 & 864.7 & 565.1 & 6997 & 9553 & 0.114 & 424.1 \\
(h) & 123.4 & 5193 & 692.8 & 911.3 & 680.8 & 877.3 & 665.7 & 7816 & 10572 & 0.120 & 561.3 \\
\hline
\end{tabulary}\end{center}
\caption{Display of the Higgs boson ($h^0$) mass, the $\mu$ parameter, the stau mass,  the relevant electroweak gaugino masses, and the relic density for the benchmarks  of Table~\ref{tab1} computed at the electroweak scale. The track length, $c\tau_0$ (in mm) left by the long-lived stau is also shown. All masses are in GeV. }
\label{tab2}
\end{table}

In Table~\ref{tab2}, all the benchmarks satisfy the Higgs boson mass and the relic density constraints. The LSP mass, as well as the masses of the MSSM neutralino $\tilde\chi^0_1$ and of the chargino $\tilde\chi^{\pm}_1$ are shown. Also the masses of the stau and tau sneutrino are given. Here the stau is the lighter of the two staus which can be made lighter than the tau sneutrino with a large off-diagonal element in the stau mass-squared matrix. The mass gap between the NLSP and the hidden sector LSP ranges from $\sim 8$ GeV (for point (f)) to $\sim 20$ GeV (for point (e)). The only decay mode of the stau is to the hidden sector neutralino, i.e. $\tilde\tau \rightarrow \tau\tilde\xi^0_1$. The smallness of the available phase space suppresses the stau decay width. Another source of suppression comes from the fact that the MSSM particles communicate with the hidden sector particles only through the small kinetic mixing coefficient $\delta$ which, according to Table~\ref{tab1}, is chosen to be very small, i.e. $\mathcal{O}(10^{-6})$. The coupling between the stau and the LSP is proportional to 
\begin{align} 
 &\frac{i}{2} \left(\sqrt{2} g_Y N^*_{1 3} \tilde{D}^{\ell}_{{13}}  +\sqrt{2} g_2 N^*_{1 4}\tilde{D}^{\ell}_{{13}}-\sqrt{2} g_Y N^*_{1 2} \tilde{D}^{\ell}_{{13}}s_{\delta}-\frac{2\sqrt{2}m_{\tau}}{v_d} N^*_{1 5} \tilde{D}^{\ell}_{{16}}\right)P_L \nonumber \\ 
  & + \,i \left[\sqrt{2} g_Y \tilde{D}^{\ell}_{{16}}\Big(- N_{{1 3}}  + N_{{1 2}}s_{\delta} \Big) - \frac{\sqrt{2}m_{\tau}}{v_d}\tilde{D}^{\ell}_{{13}} N_{{1 5}} \right]P_R,
  \label{coupling}
\end{align}
where $\tilde{D}^{\ell}$ is the matrix that diagonalizes the $6\times 6$ slepton mass-squared matrix,
\begin{equation}
\tilde{D}^{\ell}M^2_{\tilde\ell}\tilde{D}^{\ell^{\dagger}}=\text{diag}(m^2_{\tilde\ell_1},m^2_{\tilde\ell_2},m^2_{\tilde\ell_3},m^2_{\tilde\ell_4},m^2_{\tilde\ell_5},m^2_{\tilde\ell_6}),
\end{equation}
and $N$ is the matrix that diagonalizes the $6\times 6$ neutralino mass matrix,
\begin{equation}
N^* M_{\tilde\chi^0}N^{\dagger}=\text{diag}(m_{\tilde\xi^0_1},m_{\tilde\xi^0_2},m_{\tilde\chi^0_1},m_{\tilde\chi^0_2},m_{\tilde\chi^0_3},m_{\tilde\chi^0_4}).
\end{equation}
Here, $P_L$ $(P_R)$ is the left (right) projection operator and $m_{\tau}$ the tau mass. 
 
The hidden sector LSP, $\tilde\xi^0_1$, is an admixture of the $U(1)_X$ gaugino $\lambda_X$, the Majorana spinor $\psi_S$, see Eq.~(\ref{spinors}), and the visible sector (MSSM) binos, winos and higgsinos, i.e. 
\begin{equation}
\tilde\xi^0_1=N_{11}\psi_S+N_{12}\lambda_X+N_{13}\lambda_Y+N_{14}\lambda_3+N_{15}\tilde h_1+N_{16}\tilde h_2.
\label{neutralino}
\end{equation}
Since the hidden sector neutralinos interact with the visible sector only minimally, the bino, wino and higgsino contents of $\tilde\xi^0_1$ are negligible, i.e. $N_{13}\approx N_{14}\approx N_{15}\approx N_{16} \approx 0$. Further, since $s_{\delta} \ll 1$, $N_{12}s_{\delta} \ll 1$ and so the coupling given by Eq.~(\ref{coupling}) is very small. This leads to a further suppression of the stau decay width. In fact, the stau decay widths for the benchmark points of Table~\ref{tab1} are $\mathcal{O}(10^{-16})$ GeV which results in a large decay length, $c\tau_0$, as shown in Table~\ref{tab2}. 

Other than their direct production, staus can be produced following the decay of a tau sneutrino. Thus, for our benchmark points of Table~\ref{tab1}, the tau sneutrino decays predominantly to a stau and a $W$ boson with branching ratios ranging from $70\%$ to $98\%$. Thus, we will also consider the production of sneutrinos which are a source of staus as well as the direct production of staus. We note that in Table~\ref{tab1}  the kinetic mixing  parameter $\delta$ is chosen in the range $\sim 10^{-5}-10^{-6}$ so that staus  decay in the inner detector tracker. Theoretically $\delta$ arises at the loop level from  mixings between the hidden sector and the visible sector. The size of the mixing depends on the model and its value can range from 
$10^{-3}$ to  orders of magnitude smaller depending on the model~\cite{Dienes:1996zr}. Values of $\delta$ in Table~\ref{tab1}
lie well within this range.
 
A comment regarding the $Z$ and $Z'$ bosons is in order. For the benchmarks of Table~\ref{tab1}, the $Z'$ mass obtained from Eq.~(\ref{bosons}) is $\sim M_1$ since $M_2 \sim 0$ and $s_{\delta} \ll 1$. Thus the spectrum contains a $Z'$ with a mass range of $\sim 420$ GeV to $\sim 700$ GeV. However, due to the very small coupling between this $Z'$ boson and the SM particles, its production cross-section at $pp$ colliders is extremely suppressed and thus such a mass range can easily escape detection and so the typical experimental bounds on the $Z'$ mass or on $m_{Z'}/g_X$ do not apply here~\cite{Tanabashi:2018oca}. According to Eq.~(\ref{bosons}), the $Z$ boson mass receives a correction due to gauge kinetic and mass mixings. Knowing that $M_2\ll M_1$ and $s_{\delta}\ll 1$, we can write $M^2_-$ as
\begin{equation}
M^2_-\simeq M^2_Z+\frac{\epsilon}{2}g^2_Yv^2 \frac{s_{\delta}}{c_{\delta}}+\frac{1}{4}g^2_2 v^2\left(\frac{\epsilon}{\kappa}\right)^2.
\end{equation}
According to Eq.~(\ref{m2rge}), $M_2$ develops a tiny value at the electroweak scale. For the benchmark points, $\epsilon$ takes values in the range $\mathcal{O}(10^{-7})$$-$$\mathcal{O}(10^{-6})$ with $\kappa\sim 1$. Such a value gives a correction of 1 part in $10^9$ for $M_Z$ which is far beyond the sensitivity of current experiments.

\section{Dark matter relic density}\label{sec:DM}

In the standard approach to calculating the dark matter relic density, the LSP is assumed to be in thermal equilibrium with the bath and has efficient self-annihilation to SM particles which will eventually deplete the relic abundance until freeze-out sets in. In SUSY models, obtaining a bino-like LSP (lightest MSSM neutralino) is usually problematic for dark matter relic density. In this case, the self-annihilation of the LSP is suppressed and one needs coannihilation to deplete the relic density to its experimentally observed value~\cite{Aghanim:2018eyx},
\begin{equation}
\Omega h^2=0.1198\pm 0.0012.
\label{relic}
\end{equation} 

In the analysis here, the LSP is not bino-like but the hidden sector neutralino, $\tilde\xi^0_1$, which has very weak couplings to the MSSM particles and so self-annihilation is extremely small. The next odd sector particle, the NLSP, is the stau, $\tilde\tau$ with
\begin{equation}
\frac{m_{\tilde\tau}-m_{\tilde\xi^0_1}}{m_{\tilde\xi^0_1}} < 10\%,
\label{deltam}
\end{equation}

so that coannihilation is generally effective. Thus, one can have three processes responsible for the observed relic density of $\tilde\xi^0_1$, namely,
\begin{align}
\tilde\xi^0_1~\tilde\xi^0_1 \rightarrow \rm SM, \nonumber \\
\tilde\xi^0_1~\tilde\tau \rightarrow \rm SM^{\prime}, \nonumber \\
\tilde\tau~\tilde\tau \rightarrow \rm SM^{\prime\prime}, 
\label{coann}
\end{align}
where SM, SM$^\prime$, SM$^{\prime\prime}$ stand for some standard model particles. To have a feel for the size of the first process in Eq.~(\ref{coann}) , i.e. LSP self-annihilation, we consider the $\tilde\xi^0_1\tilde\xi^0_1 Z$ vertex, whose coupling is proportional to
\begin{equation}
\frac{ig_2}{2\cos\theta_W}\gamma^{\mu}(\sin\theta_W\sin\theta'_W s_{\delta}+\cos\theta'_W)(|N_{15}|^2-|N_{16}|^2)\gamma_5,
\end{equation}
where $|N_{15}|^2$ and $|N_{16}|^2$ represent the higgsino content of $\tilde\xi^0_1$ [see Eq.~(\ref{neutralino})] which is negligible due to the very weak interaction of the LSP with the MSSM particles. Hence the annihilation cross-section of two LSPs is found to be extremely small. The second process of Eq.~(\ref{coann}) is inefficient as well due to the smallness of the coupling from Eq.~(\ref{coupling}) which is $\mathcal{O}(10^{-6})$. The only channel with efficient annihilation is the last one which involves $\tilde\tau\tilde\tau$ with purely MSSM interactions and no dependence on $s_{\delta}$ but has a larger Boltzmann suppression $\sim e^{-2m_{\tilde\tau}/T}$ and thus the reason for condition Eq.~(\ref{deltam}). For such very weak couplings of the dark matter particle, one should ask whether chemical equilibrium can be achieved, i.e.,
\begin{equation}
\frac{n_{\tilde\xi^0_1}}{n_{\tilde\tau}}=\frac{n^{\rm eq}_{\tilde\xi^0_1}}{n^{\rm eq}_{\tilde\tau}},
\label{CE}
\end{equation}
where $n$ is the number density and $n^{\rm eq}$ is the equilibrium number density. Chemical equilibrium is generally guaranteed if conversion-driven processes such as co-scattering $\tilde\xi^0_1~\rm SM\leftrightarrow\tilde\tau ~\rm SM$, decay and inverse decay of the NLSP, $\tilde\xi^0_1~\rm SM\leftrightarrow\tilde\tau$ are fast enough around the time of freeze-out. In this case one must solve the coupled Boltzmann equations~\cite{Garny:2017rxs,Garny:2018icg} which include those conversion-driven processes. The full coupled set of Boltzmann equations pertaining to $\tilde\xi^0_1$ and $\tilde\tau$ are given below in Eqs.~(\ref{boltzmann1}) and~(\ref{boltzmann2}), which take into consideration all conversion-driven processes (LSP $\leftrightarrow$ NLSP), 
\begin{align}
\frac{dY_{\tilde\xi^0_1}}{dx}&=\frac{1}{3H}\frac{ds}{dx}\Bigg[\langle \sigma_{\tilde\xi^0_1\tilde\xi^0_1} v\rangle(Y^2_{\tilde\xi^0_1}-Y^{\rm eq 2}_{\tilde\xi^0_1})+\langle \sigma_{\tilde\xi^0_1\tilde\tau} v\rangle(Y_{\tilde\xi^0_1}Y_{\tilde\tau}-Y^{\rm eq}_{\tilde\xi^0_1}Y^{\rm eq}_{\tilde\tau})+\frac{\Gamma_{\tilde\xi^0_1\rm SM\rightarrow\tilde\tau \rm SM}}{s}\left(Y_{\tilde\xi^0_1}-Y_{\tilde\tau}\frac{Y^{\rm eq}_{\tilde\xi^0_1}}{Y^{\rm eq}_{\tilde\tau}}\right) \nonumber \\
&-\frac{\Gamma_{\tilde\xi^0_1\rm SM\leftrightarrow\tilde\tau}}{s}\left(Y_{\tilde\tau}-Y_{\tilde\xi^0_1}\frac{Y^{\rm eq}_{\tilde\tau}}{Y^{\rm eq}_{\tilde\xi^0_1}}\right)+\langle \sigma_{\tilde\xi^0_1\tilde\xi^0_1\rightarrow\tilde\tau^+\tilde\tau^-} v\rangle\left(Y^2_{\tilde\xi^0_1}-Y^2_{\tilde\tau}\frac{Y^{\rm eq2}_{\tilde\xi^0_1}}{Y^{\rm eq2}_{\tilde\tau}}\right)\Bigg],
\label{boltzmann1}
\end{align}
and
\begin{align}
\frac{dY_{\tilde\tau}}{dx}&=\frac{1}{3H}\frac{ds}{dx}\Bigg[\langle \sigma_{\tilde\tau^+\tilde\tau^-} v\rangle(Y^2_{\tilde\tau}-Y^{\rm eq 2}_{\tilde\tau})+\langle \sigma_{\tilde\xi^0_1\tilde\tau} v\rangle(Y_{\tilde\xi^0_1}Y_{\tilde\tau}-Y^{\rm eq}_{\tilde\xi^0_1}Y^{\rm eq}_{\tilde\tau})-\frac{\Gamma_{\tilde\xi^0_1\rm SM\rightarrow\tilde\tau \rm SM}}{s}\left(Y_{\tilde\xi^0_1}-Y_{\tilde\tau}\frac{Y^{\rm eq}_{\tilde\xi^0_1}}{Y^{\rm eq}_{\tilde\tau}}\right) \nonumber \\
&+\frac{\Gamma_{\tilde\xi^0_1\rm SM\leftrightarrow\tilde\tau}}{s}\left(Y_{\tilde\tau}-Y_{\tilde\xi^0_1}\frac{Y^{\rm eq}_{\tilde\tau}}{Y^{\rm eq}_{\tilde\xi^0_1}}\right)-\langle \sigma_{\tilde\xi^0_1\tilde\xi^0_1\rightarrow\tilde\tau^+\tilde\tau^-} v\rangle\left(Y^2_{\tilde\xi^0_1}-Y^2_{\tilde\tau}\frac{Y^{\rm eq2}_{\tilde\xi^0_1}}{Y^{\rm eq2}_{\tilde\tau}}\right)\Bigg],
\label{boltzmann2}
\end{align} 

with $Y=n/s$, where $s$ is the entropy density and $x=m_{\tilde\xi^0_1}/T$. The first two terms in Eq.~(\ref{boltzmann1}) are negligible and so is the second term in Eq.~(\ref{boltzmann2}). The last three terms in each of those equations represent the conversion terms which play an important role in establishing DM freeze-out. The last term in Eqs.~(\ref{boltzmann1}) and~(\ref{boltzmann2}) represents scattering of DM particles into odd sector particles, $\tilde\xi^0_1\tilde\xi^0_1\rightarrow\tilde\tau^+\tilde\tau^-$ (last term) which is negligible due to the very weak coupling of $\tilde\xi^0_1$ and the thermal suppression by $n^{\rm eq}_{\tilde\xi^0_1}$. \\
Since $m_{\tilde\tau}>m_{\tilde\xi^0_1}$ the co-scattering of $\tilde\xi^0_1$ into $\tilde\tau$ requires that $\tilde\xi^0_1$ and the SM particle have enough momentum, such a process is highly momentum-dependent. However, for the benchmark points of Table~\ref{tab1}, the stau can decay into an LSP and a tau such that $m_{\tilde\tau}>m_{\tilde\xi^0_1}+m_{\tau}$ and the decay width of the stau, $\Gamma_{\tilde\tau}$, ranges from $\sim 3.5\times 10^{-16}$ GeV to $\sim 1.3\times 10^{-15}$ GeV. Knowing that the Hubble parameter, $H(T)$ is given at a temperature $T$ by
\begin{equation}
H(T)=\sqrt{\frac{4\pi^3g_* G_N}{45}}T^2 \sim 10^{-18}~T^2,
\end{equation} 
it is found that $H(T_f)<\Gamma_{\tilde\tau}$ for a freeze-out temperature $T_f=m_{\tilde\xi^0_1}/x_f$ for the benchmark points (a)$-$(f), where the average freeze-out temperature occurs for $x_f\sim 26.5$. The forward-backward processes $\tilde\tau \leftrightarrow \tilde\xi^0_1 \tau$ help equilibrate $\tilde\xi^0_1$ and $\tilde\tau$. Notice that the inverse decay plays the same role as co-scattering but has a larger rate. The conversion of $\tilde\xi^0_1$ into a $\tilde\tau$ is followed by stau self-annihilation into SM particles via the dominant processes $\tilde\tau^+\tilde\tau^-\rightarrow h^0h^0$ and $\tilde\tau^+\tilde\tau^-\rightarrow W^+W^-$ which eventually deplete the relic abundance satisfying the current limit as shown in Table~\ref{tab2}. So in principle since the inverse decay channel is open, the relic density is determined by coannihilation because inverse decay processes decouple later. For points (g) and (h), $H(T_f)>\Gamma_{\tilde\tau}$ and so the process $\tilde\tau \leftrightarrow \tilde\xi^0_1 \tau$ decouples which is when co-scattering $\tilde\xi^0_1~ \rm SM \leftrightarrow \tilde\tau~ \rm SM'$ starts playing an important role in converting DM particles into $\tilde\tau$ followed by annihilation into SM particles, $\tilde\tau\tilde\tau\rightarrow \rm SM~SM$.

To summarize, the hidden sector communicates with the MSSM via the kinetic mixing coefficient and for $s_{\delta} \gtrsim 10^{-6}$ the dark sector is in kinetic equilibrium with the MSSM~\cite{DAgnolo:2017dbv}. For the coupling strengths considered in this analysis, the DM particle annihilation and coannihilation via $\tilde\xi^0_1\tilde\xi^0_1\rightarrow \rm SM ~SM$ and $\tilde\xi^0_1\tilde\tau \rightarrow \rm SM ~SM$ are negligible whereas $\tilde\tau\tilde\tau \rightarrow \rm SM ~SM$ is dominant. For fast decay and inverse decay of $\tilde\tau$ (which sets the chemical equilibrium between $\tilde\xi^0_1$ and $\tilde\tau$), co-scattering processes do not contribute to the relic density and the latter is merely determined by coannihilation (i.e. by $\tilde\tau$ self-annihilation)~\cite{Filimonova:2018qdc}. When the decay width of $\tilde\tau$ falls below the Hubble parameter around freeze-out, coannihilation and co-scattering freeze-out will determine the final relic abundance~\cite{Cheng:2018vaj}. Here, two cases arise: if the freeze-out temperature of coannihilation is larger than co-scattering then the former freezes out earlier thus the number of $\tilde\xi^0_1$ and $\tilde\tau$ in a comoving volume is fixed. Co-scattering processes only redistribute the two particles' number densities and so the relic density is set by coannihilation. If the freeze-out temperature of co-scattering is greater than coannihilation then co-scattering freezes-out first which means the LSP is no longer being converted to the NLSP. The remaining NLSPs will be removed by coannihilation (or self-annihilation to be precise). Therefore the relic density is set by co-scattering.

For even weaker couplings of the dark sector, the LSP may fall out of thermal equilibrium and decouple from the bath soon after being produced. Such a particle is known as a FIMP (feebly interacting massive particle). If in the early universe, the LSPs had little initial abundance due to inflationary effects or other mechanisms then even though the interactions with the bath is feeble, dark matter particles may still be produced over time until the interaction rate falls below the expansion rate of the universe and the relic abundance ``freezes-in". This is known as the freeze-in mechanism~\cite{Hall:2009bx,Belanger:2018mqt} which can be viewed as the opposite of the usual freeze-out mechanism where one starts with a huge initial abundance of dark matter particles which are in thermal equilibrium with the bath. The production of such a feeble particle is through the decays of heavier particles. For a certain range of couplings, the LSP relic density can be even due to both contributions from freeze-out and freeze-in~\cite{Banerjee:2018uut}. For the range of couplings we consider, freeze-in does not factor in and the relic abundance is purely due to the freeze-out of the LSP via the mechanisms described above.

\section{Stau pair production and stau associated production with a sneutrino at the LHC}\label{sec:production}

The main mechanism for the production of a light stau at the LHC is through pair production, $pp\rightarrow \tilde\tau^{+}\tilde\tau^{-}$ and associated production with a tau sneutrino, $pp\rightarrow \tilde\tau\tilde\nu_{\tau}$. In the $U(1)_X$-extended MSSM/SUGRA, the stau pair production proceeds via $\gamma$, $Z$ and $Z'$ $s$-channel processes, i.e. $q\bar{q}\rightarrow \gamma,Z,Z'\rightarrow \tilde\tau^{+}\tilde\tau^{-}$, whereas stau associated production with a tau sneutrino proceeds by the exchange of $W^{\pm}$ boson. The coupling of $Z'$ to fermions is small and in particular the coupling to up-type quarks is proportional to   
\begin{equation}
-ig_2\frac{\sin\theta'_W}{\cos\theta_W}\gamma^{\mu}\frac{1}{2}\left[-\frac{1}{2}+\frac{4}{3}\sin^2\theta_W-\frac{5}{6}\sin\theta_W s_{\delta}\cot\theta'_W+\frac{1}{2}\left(1-\sin\theta_W s_{\delta}\cot\theta'_W\right)\gamma_5\right].
\label{zpc}
\end{equation}
Since $\epsilon\ll 1$ the mixing angle $\theta'_W$ is very small and so is $s_{\delta}$ which means that the coupling of Eq.~(\ref{zpc}) is small as well. For this reason, the contribution to the cross-section from $Z'$ can be neglected. Thus, the production cross-section of the stau pair can be determined directly from the MSSM. We calculate the di-stau and stau-tau sneutrino LHC production cross-sections using \code{Prospino2}~\cite{Beenakker:1996ed,Beenakker:1999xh} at the next-to-leading order (NLO) in QCD at 14 TeV and at 27 TeV using the CTEQ5 PDF set~\cite{Lai:1999wy}. The results of the analysis are presented in Table~\ref{tab3}. Note that for a $pp$ collider the production cross-section of $\tilde\tau^+\tilde\nu_{\tau}$ is larger than $\tilde\tau^-\tilde\nu^*_{\tau}$.   

\begin{table}[H]
\begin{center}
\begin{tabulary}{1.12\textwidth}{l|cc|cc|cc}
\hline\hline\rule{0pt}{3ex}
Model  & \multicolumn{2}{c}{$\sigma_{\rm NLO}(pp\rightarrow \tilde\tau^{+}\tilde\tau^{-})$} & \multicolumn{2}{c}{$\sigma_{\rm NLO}(pp\rightarrow \tilde\tau^+\tilde\nu_{\tau})$} & \multicolumn{2}{c}{$\sigma_{\rm NLO}(pp\rightarrow \tilde\tau^-\tilde\nu^*_{\tau})$} \\
&  &  &  &  &  & \\ 
  & 14 TeV & 27 TeV & 14 TeV & 27 TeV & 14 TeV & 27 TeV \\ 
\hline\rule{0pt}{3ex} 
\!\!(a) & 2.70 & 8.05 & 2.05 & 6.45 & 0.90  & 3.42  \\
(b) & 2.03 & 6.17 & 0.48 & 1.67 &  0.19 & 0.84  \\
(c) & 1.74 & 5.53 & 1.94 & 6.39 & 0.83  &  3.30 \\
(d) & 0.41 & 1.50 & 0.17 & 0.68 & 0.06 & 0.32  \\
(e) & 0.49 & 1.85 & 0.74 & 2.84 &  0.29 & 1.37  \\
(f) & 0.21 & 0.85 & 0.21 & 0.90 & 0.08  & 0.42  \\
(g) & 0.10 & 0.45 & 0.05 & 0.24 & 0.02 & 0.11  \\
(h) & 0.04 & 0.25 & 0.06 & 0.32 &  0.02 &  0.13 \\
\hline
\end{tabulary}
\end{center}
\caption{The NLO production cross-sections, in fb, of a stau pair, $\tilde\tau^+\tilde\tau^-$ (second and third columns), and $\tilde\tau\tilde\nu_{\tau}$ (fourth, fifth, sixth and seventh columns), at $\sqrt{s}=14$ TeV and at $\sqrt{s}=27$ TeV for benchmarks of Table~\ref{tab1}.}
\label{tab3}
\end{table}

\section{Signal and background simulation and event selection}\label{sec:simulation}

\begin{figure}[H]
 \centering
 	\includegraphics[width=0.7\textwidth]{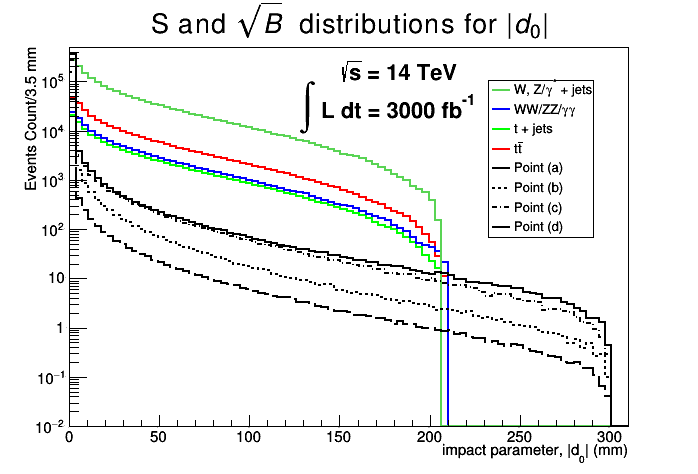}
      \caption{A distribution of the impact parameter, $|d_0|$, for the parameter points (a), (b), (c), (d) of Table~\ref{tab1} and the SM background at 14 TeV and 3000$\ifb$ of integrated luminosity. No selection criteria have been imposed in this distribution.}
	\label{fig1}
\end{figure}

Our signal consists of a mixture of stau pair production and stau associated production with a tau sneutrino. The end products of the decay chain and the relevant final states are as in Eq.~(\ref{finalstates})
\begin{align}
pp\rightarrow \tilde\tau^+\tilde\tau^-\rightarrow \tau^+\tau^-\tilde\xi^0_1\tilde\xi^0_1\rightarrow \tau_h,\ell+E^{\rm miss}_T, \nonumber  \\
pp\rightarrow \tilde\tau^{\pm}\tilde\nu_{\tau}\rightarrow \tau^{\pm}\tilde\xi^0_1\tau^{\pm}W^{\mp}\rightarrow \tau_h,2\ell+E^{\rm miss}_T,
\label{finalstates}
\end{align}
where $\tau_h$ corresponds to a hadronically decaying $\tau$, $\ell$ represents a light lepton (electron or muon) and $E^{\rm miss}_T$ is the missing transverse energy due to neutrinos and the LSP. The event preselection criteria involves at least one isolated light lepton and at most one hadronically decaying tau to retain as much signal as possible. No selection criteria is imposed on the missing transverse energy, $E^{\rm miss}_T$, as in most of the parameter points considered the final states involve little $E^{\rm miss}_T$ which in most situations is below the dectectors' trigger level. Furthermore, since our stau is long-lived, it will leave a track in the inner detector (ID) tracker characterized by low speed and large invariant mass. We are interested in looking at tracks left by charged particles (mostly leptons) originating from the decay of the long-lived stau. Some studies already exist in this direction, see, e.g.~\cite{Evans:2016zau}. Since the lepton track is soft (of low $p_T$), the combination of the stau track and lepton track constitute what is known as a kinked track~\cite{Lee:2018pag}. The lepton tracks are highly displaced and so are characterized by a large impact parameter, $d_0$, which is the shortest distance, in the $(x,y)$ plane perpendicular to the beams' direction, between the track and the collision point. Such a signature is a combination between kinked and displaced tracks. 
The decay length of the long-lived stau can be determined by
\begin{equation}
d_{xy}=\sqrt{(x_m-x_p)^2+(y_m-y_p)^2},
\end{equation}     
where $(x_m,y_m)$ and $(x_p,y_p)$ are the vertex coordinates of the mother and daughter particles, respectively. For the long-lived stau, $(x_p,y_p)$ represents the vertex (or track initial point) of the tau (or the resulting leptons) with a large impact parameter and $(x_m,y_m)$ is taken to be $(0,0)$, i.e. at the primary vertex. It is known that imposing cuts on the impact parameter and decay length as $|d_0|>($2$-$4) mm and $d_{xy}>($4$-$8) mm will greatly reduce the SM background~\cite{Chatrchyan:2012jna,Aad:2012zx,Cerdeno:2013oya,ATLAS:2012av,Abdallah:2018gjj}. \\
Given the final states of Eq.~(\ref{finalstates}), the largest contributors to the physical SM backgrounds are $W/Z/\gamma^*+$ jets, diboson production, single top and $t\bar{t}$. The signal and background events are simulated at leading order (LO) with \code{MadGraph5\_aMC@NLO-2.6.3} interfaced to \code{LHAPDF}~\cite{Buckley:2014ana} using the NNPDF30LO PDF set. The cross-sections are then scaled to their NLO values at 14 TeV and at 27 TeV. The resulting files are passed to \code{PYTHIA8}~\cite{Sjostrand:2014zea} for showering and hadronization. For the SM backgrounds, a five-flavour MLM matching~\cite{Mangano:2006rw} is performed on the samples in order to avoid double counting of jets. Jets are clustered with \code{FastJet}~\cite{Cacciari:2011ma} using the anti-$k_t$ algorithm~\cite{Cacciari:2008gp} with jet radius $R=0.4$. Detector simulation and event reconstruction is handled by \code{DELPHES-3.4.2}~\cite{deFavereau:2013fsa} using the beta card for HL-LHC and HE-LHC studies. The analysis of the resulting event files and cut implementation is carried out with \code{ROOT 6}~\cite{Antcheva:2011zz}. The unphysical background contamination is due to fake tracks arising from the high pile-up environment. We simulate minimum bias events due to elastic and inelastic (diffractive and non-diffractive) soft QCD events with \code{PYTHIA8} which are mixed with the main interaction. We consider a mean pile-up (interactions per bunch crossing) of 128~\cite{ATLAS:2013hta} for both HL-LHC and HE-LHC\footnote{Estimated pile-up at the HL-LHC may reach $\sim 200$ while at HE-LHC the figure may rise up to $\sim 800$.}. Pile-up mitigation is handled by \code{PUPPI}~\cite{Bertolini:2014bba} with the default settings used for CMS phase II Delphes card. \\
To show the size of the impact parameter of the signal events in relation to the SM background, we present such a distribution in Fig.~\ref{fig1}. The benchmarks (a), (b), (c) and (d) are shown by the black histograms whereas the SM backgrounds are represented by the colored ones. Note that no preselection cuts have been imposed yet on the signal and background in this plot. One can clearly see that the SM background events fall to zero at $|d_0|\sim 200$ mm whereas signal events extend all the way up to $\sim 300$ mm as a result of the late decay products of the stau.

\section{Cut-flow analysis and results}\label{sec:results}

We give a cut-and-count analysis for the discovery potential of a long-lived stau for the signal benchmark points of Table~\ref{tab1} at the HL-LHC and HE-LHC by comparing results of the number of signal events surviving the cuts at select integrated luminosities due to zero and non-zero pile-up environments. Even though we know that at high luminosities pile-up will be a significant player, an analysis with no pile-up would give us an idea on how the performance is affected by adding pile-up which is a sign of the effectiveness of the considered pile-up subtraction algorithm.  \\
As explained in the section~\ref{sec:simulation}, we are looking for a light lepton (electron or muon) track with high impact parameter, $|d_0|$, originating from a high momentum track due to the long-lived charged stau. Hereafter, we list the kinematic variables used to discriminate the signal from the SM background:
\begin{enumerate}
\item $|d_0|$: the track impact parameter which is chosen to be large enough to eliminate as many background events as possible.
\item $p_T^{e~[\mu]}$: the transverse momentum of an isolated electron or muon. 
\item $p_T^{\rm tracks}$: the transverse momentum of tracks in the ID.
\item Isolated lepton tracks: the number of isolated leptons must match the number of lepton tracks. This ensures that we reject any lepton tracks which are not isolated. 
\item $\Delta R(\tilde\tau,\rm track)$: the minimum spatial separation between the lepton tracks and the stau track. A small cut on this variable ensures that the lepton track considered has originated from a long-lived stau.
\item $\beta=p/E$: the velocity of the long-lived particle. A cut on $\beta$ allows us to reject events with muons faking a stau track.   
\end{enumerate}

\begin{figure}[H]
\centering
\includegraphics[width=0.49\textwidth]{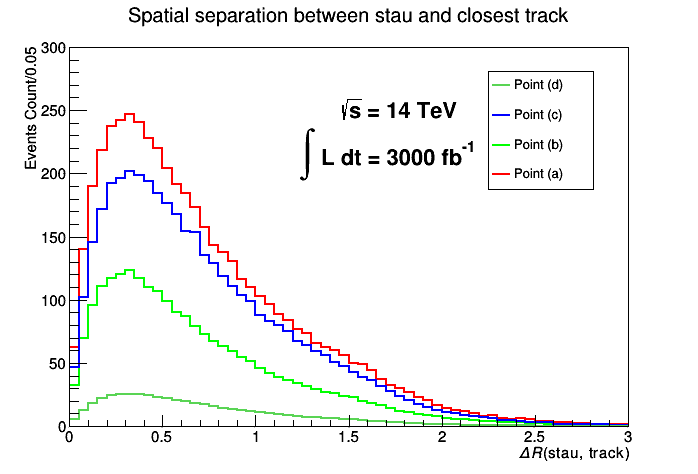}
\includegraphics[width=0.49\textwidth]{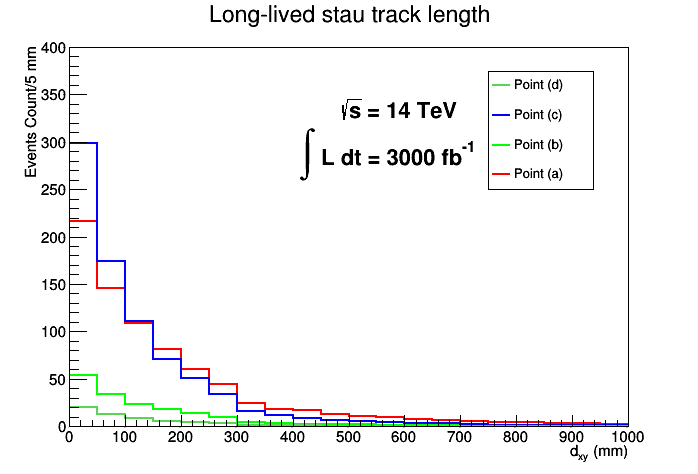}
\caption{Left panel: Minimum spatial separation between the stau LLP and its closest lepton track, $\Delta R(\tilde\tau,\rm track)$. Right panel: the track length $d_{xy}$, of the long-lived stau.}
\label{fig2}
\end{figure}

We present in the left panel of Fig.~\ref{fig2} $\Delta R(\tilde\tau,\rm track)$ which has peak values for small spatial separation. Thus a cut of $\Delta R(\tilde\tau,\rm track)<0.6$ should be sufficient to ensure that the lepton tracks have actually originated from the corresponding stau track. The right panel of Fig.~\ref{fig2} displays the decay length, $d_{xy}$ of the stau which clearly can travel up to 1 m in the ID, knowing that the typical tracker radius is between 35 mm and 1200 mm.

\subsection{Results with no pile-up}\label{sec:noPU}
We start by showing results for the case of no pile-up. After applying the preselection cuts, cuts on the kinematic variables 1$-$6 of section~\ref{sec:results} are applied on the signal and background samples. We give in Table~\ref{tab4} the cut-flow for three parameter points, (a), (c) and (f) and the SM backgrounds where the samples are normalized to their cross-section values, in fb. The points are chosen to represent cases of maximal (c), moderate (a) and low (f) signal event yield. It is clear that no backgrounds survive the cuts as one would expect from such a signal topology. Note that the two kinematic variables that have the most impact on the backgrounds (and partly on the signal) are $|d_0|$ and the track isolation condition. In an actual collider experiment, the backgrounds are not exactly zero but are mostly instrumental in nature~\cite{Aaboud:2017iio}. The other background sources come from accidental crossing of tracks especially in pile-up environments. We will consider this when discussing pile-up in our analysis next section. 

\begin{table}[H]
\begin{center}
\resizebox{\linewidth}{!}{\begin{tabulary}{0.85\textwidth}{l|ccccccc}
\hline\hline\rule{0pt}{3ex}
Cuts & (a) & (c) & (f) & $t\bar{t}$ & $t+$jets & $W/Z/\gamma^*+$ jets&  $WW/ZZ/\gamma\gamma$ \\
      \hline
$N(\ell)\geq 1$ & 1.61 & 1.35 & 0.0795 & 236731 & 38618 & $4.39\times 10^6$ & 79831  \\
$N(\tau_h)\leq 1$ & 1.60 & 1.34 & 0.0794 & 236126 & 38596 & $4.38\times 10^6$ &  79796 \\
$|d_0|>4$ mm & 0.27 & 0.29 & 0.04 & 1803 & 190 & 6063 & 101 \\
$p_T^{e~[\mu]}>15~[10]$ GeV & 0.084 & 0.096 & 0.0093 & 1494 & 142 & 4871 & 81 \\
$p_T^{\rm tracks}>50$ GeV & 0.050 & 0.061 & 0.0036 & 1168 & 103 & 2902 & 48 \\
Isolated lepton tracks & 0.016 & 0.021 & 0.00057 & 1.02 & 0.09 & 0 & 0 \\
$d_{xy}>20$ mm & 0.015 & 0.0197 & 0.00055 & 0 & 0 & 0 & 0 \\
$\Delta R(\tilde\tau,\rm track)<0.6$ & 0.011 & 0.013 & 0.00042 & 0 & 0 & 0 & 0 \\
$\beta<0.95$ & 0.0093 & 0.012 & 0.00040 & 0 & 0 & 0 & 0 \\
\hline
\end{tabulary}}\end{center}
\caption{Cut-flow for parameter points (a), (c) and (f) and SM background at $\sqrt{s}=14$ TeV for the case of no pile-up. Samples are normalized to their respective cross-sections (in fb).}
\label{tab4}
\end{table}

For the eight benchmark points of Table~\ref{tab1} we give the projected number of signal events surviving the cuts at select integrated luminosities at the HL-LHC. We present the results in Table~\ref{tab5}. One can see that four out of the eight points may be discovered at the HL-LHC with integrated luminosities up to $3000 \ifb$ where we are assuming that a signal event yield $>5$ is enough to claim discovery over an almost zero background.  

\begin{table}[H]
\begin{center}
\begin{tabulary}{2.15\textwidth}{l|ccccc}
\hline\hline\rule{0pt}{3ex}
Model & $\mathcal{N}_{\rm events}^{500\ifb}$ & $\mathcal{N}_{\rm events}^{1000\ifb}$ & $\mathcal{N}_{\rm events}^{1500\ifb}$ & $\mathcal{N}_{\rm events}^{2000\ifb}$ \\
      \hline
(a) & 4.7 & 9.3 & 14.0 & 18.6 \\
(b) & 1.4 & 2.8 & 4.2 & 5.7 \\
(c) & 6.0 & 12.0 & 18.0 & 24.0 \\
(d) & $<1$ & $<1$ & 1.0 & 1.4 \\
(e) & 2.2 & 4.3 & 6.5 & 8.6 \\
(f) & $<1$ & $<1$ & $<1$ & $<1$ \\
(g) & $<1$ & $<1$ & $<1$ & $<1$ \\
(h) & $<1$ & $<1$ & $<1$ & $<1$ \\
\hline
\end{tabulary}\end{center}
\caption{Projected number of signal events at select integrated luminosities for benchmark points of Table~\ref{tab1} at HL-LHC for the case of no pile-up.}
\label{tab5}
\end{table}

While still in the case of no pile-up, we give the same cut-flow for the signal points (a), (c) and (f) and the SM backgrounds but for the HE-LHC. The effect of the kinematic variables on the signal and background is the same for HE-LHC as in the HL-LHC. The results are given in Table~\ref{tab6}. 

\begin{table}[H]
\begin{center}
\resizebox{\linewidth}{!}{\begin{tabulary}{0.85\textwidth}{l|ccccccc}
\hline\hline\rule{0pt}{3ex}
Cuts & Signal (a) & Signal (c) & Signal (f) & $t\bar{t}$ & $t+$jets & $W/Z/\gamma^*+$ jets&  $WW/ZZ/\gamma\gamma$ \\
      \hline
$N(\ell)\geq 1$ & 4.90 & 4.36 & 0.33 & 884189 & 111761 & $8.28\times 10^6$ & 145077  \\
$N(\tau_h)\leq 1$ & 4.88 & 4.33 & 0.32 & 881736 & 111689 & $8.27\times 10^6$ & 144975 \\
$|d_0|>4$ mm & 0.70 & 0.80 & 0.152 & 6579 & 546 & 11851 & 188 \\
$p_T^{e~[\mu]}>15~[10]$ GeV & 0.22 & 0.27 & 0.036 & 5491 & 407 & 9671 & 162 \\
$p_T^{\rm tracks}>50$ GeV & 0.13 & 0.16 & 0.014 & 4225 & 296 & 5392 & 111 \\
Isolated lepton tracks & 0.04 & 0.057 & 0.0024 & 1.90 & 0 & 0 & 0 \\
$d_{xy}>20$ mm & 0.038 & 0.053 & 0.0023 & 0 & 0 & 0 & 0 \\
$\Delta R(\tilde\tau,\rm track)<0.6$ & 0.027 & 0.037 & 0.0018 & 0 & 0 & 0 & 0 \\
$\beta<0.95$ & 0.021 & 0.029 & 0.0015 & 0 & 0 & 0 & 0 \\
\hline
\end{tabulary}}\end{center}
\caption{Cut-flow for parameter points (a), (c) and (f) and SM background at $\sqrt{s}=27$ TeV for the case of no pile-up. Samples are normalized to their respective cross-sections (in fb).}
\label{tab6}
\end{table}

One can see that at the HE-LHC all of the eight benchmarks can be discovered with integrated luminosities up to $6000 \ifb$ (see Table~\ref{tab7}).

\begin{table}[H]
\begin{center}
\begin{tabulary}{2.15\textwidth}{l|ccccccc}
\hline\hline\rule{0pt}{3ex}
Model & $\mathcal{N}_{\rm events}^{200\ifb}$ & $\mathcal{N}_{\rm events}^{300\ifb}$ & $\mathcal{N}_{\rm events}^{800\ifb}$ & $\mathcal{N}_{\rm events}^{2000\ifb}$ & $\mathcal{N}_{\rm events}^{4000\ifb}$ & $\mathcal{N}_{\rm events}^{6000\ifb}$ \\
      \hline
(a) & 4.3 & 6.4 & 17.0 & 42.6 & 85.2 & 127.8 \\
(b) & 1.3 & 1.9 & 5.2 & 13.0 & 26.0 & 39.0 \\
(c) & 5.8 & 8.7 & 23.1 & 57.7 & 115.5 & 173.2 \\
(d) & $<1$ & $<1$ & 2.2 & 5.4 & 10.9 & 16.3 \\
(e) & 3.6 & 5.4 & 14.4 & 35.9 & 71.8 & 107.6 \\
(f) & $<1$ & $<1$ & 1.2 & 3.1 & 6.2 & 9.2 \\
(g) & $<1$ & $<1$ & $<1$ & 1.7 & 3.4 & 5.0 \\
(h) & $<1$ & $<1$ & $<1$ & 1.9 & 3.8 & 5.7 \\
\hline
\end{tabulary}\end{center}
\caption{Projected number of signal events at select integrated luminosities for benchmark points of Table~\ref{tab1} at HE-LHC for the case of no pile-up.}
\label{tab7}
\end{table}

\subsection{Effect of pile-up}\label{sec:PU}

We study the effect of pile-up on the signal and background events yield by considering an average of 128 interactions per bunch crossing. The presence of pile-up increases the track multiplicity and jet activity especially in the low momentum regime. The \code{PUPPI} algorithm is used for pile-up subtraction which is based on identifying charged particles from pile-up and assigning weights for neutral ones. The weights are then used to rescale the particles' four-momenta. Hence \code{PUPPI} improves the reconstruction of objects such as jets at the particle level before the clustering sequence is initiated. Improvements have been shown also at the level of $E^{\rm miss}_T$. For this reason, we will use the \code{PUPPI} jets and $E^{\rm miss}_T$ in our kinematic variables. Thus, the same kinematic variables mentioned in section~\ref{sec:noPU} will be used here with slight modifications and additions. As can be seen from Fig.~\ref{fig3}, due to pile-up, the lepton track multiplicity has increased dramatically and the number of lepton tracks matching the number of isolated leptons (isolated lepton tracks criterion) is now very small. Applying this criterion leads to almost a loss of the entire signal while keeping a lot of background events. To mitigate this issue, we apply an additional cut on the lepton tracks thus requiring $p^{\rm tracks}_{T_{\ell}}> 5$ GeV. This cut tends to clean low momentum lepton tracks and restores the importance of the ``isolated lepton tracks" criterion.       

\begin{figure}[H]
\centering
\includegraphics[width=0.6\textwidth]{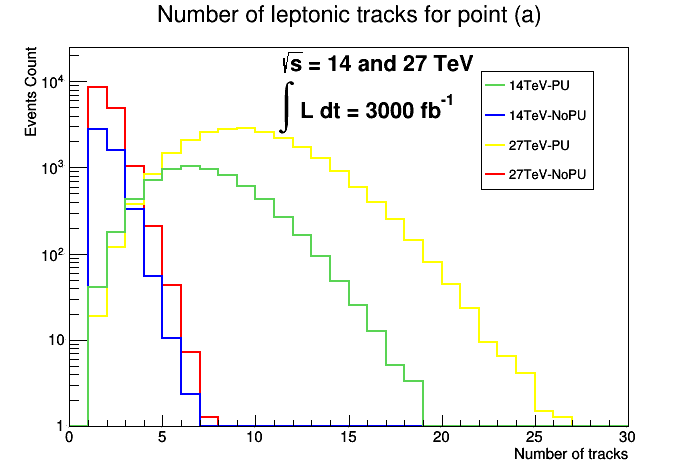}
\caption{A comparison between the number of leptonic tracks for the cases of no pile-up (NoPU) and pile-up (PU) at 14 TeV and at 27 TeV for point (a).}
\label{fig3}
\end{figure}

An additional kinematic variable, $E^{\rm miss, PUPPI}_T/\sqrt{H^{\rm PUPPI}_T}$, is used to eliminate multi-jet events that could have originated due to pile-up, where $E^{\rm miss, PUPPI}_T$ is the missing transverse energy object created after pile-up mitigation and $H^{\rm PUPPI}_T=\sum p^{\rm Hadronic}_T$ calculated using \code{PUPPI} jets. Furthermore, the cut values on some of the kinematic variables need to be adjusted to accommodate the pile-up environment. Hence harder cuts need to be applied to further clean the effects of pile-up. Different combinations of cut values were tried and the ones giving the optimal results are summarized in Table~\ref{tab8}. 

\begin{table}[H]
\begin{center}
\begin{tabulary}{2.15\textwidth}{l|cc}
\hline\hline\rule{0pt}{3ex}
Cut & 14 TeV & 27 TeV \\
      \hline
$|d_0|$ [mm] & $>8$  &  $>20$  \\
$p_T^{\rm tracks}$ [GeV] & $>50$    & $>90$    \\
$d_{xy}$ [mm] & $>20$   &  $>80$  \\
$E^{\rm miss, PUPPI}_T/\sqrt{H^{\rm PUPPI}_T}$ [GeV$^{-1/2}$ ] & $>12$   &  $>6$  \\
\hline
\end{tabulary}\end{center}
\caption{The top three are modification of the cuts given in Tables~\ref{tab4} and~\ref{tab6}, while the bottom cut is additional and used on signal and background after inclusion of pile-up. The other cuts are the same as in Tables~\ref{tab4} and~\ref{tab6}.}
\label{tab8}
\end{table}

Applying the new cut values, the signal event yield drops as one would expect in the case of pile-up. The number of signal events surviving the cuts for cases of pile-up and no pile-up are displayed in Fig.~\ref{fig4} at 14 TeV and at 27 TeV as a function of the integrated luminosity. For the 14 TeV case, only the observable points are displayed (all points are observable in the 27 TeV case). It is seen that the event yield has dropped by values ranging from $\sim 16\%$ for point (a) to $\sim 30\%$ for point (e) at 14 TeV and from $\sim 6\%$ for point (f) to $\sim 32\%$ for point (e) at 27 TeV. The experimental collaborations' techniques for pile-up mitigation is always being refined and more innovative and cutting edge tools appear regularly to try and subtract pile-up which involve many machine learning algorithms. It would be naive to linearly scale our percentage drop in yield to higher pile-up values as those results are only pertinent to the use of the \code{PUPPI} algorithm which may or may not be the algorithm of choice at the HL-LHC and HE-LHC.

\begin{figure}[H]
\centering
\includegraphics[width=0.48\textwidth]{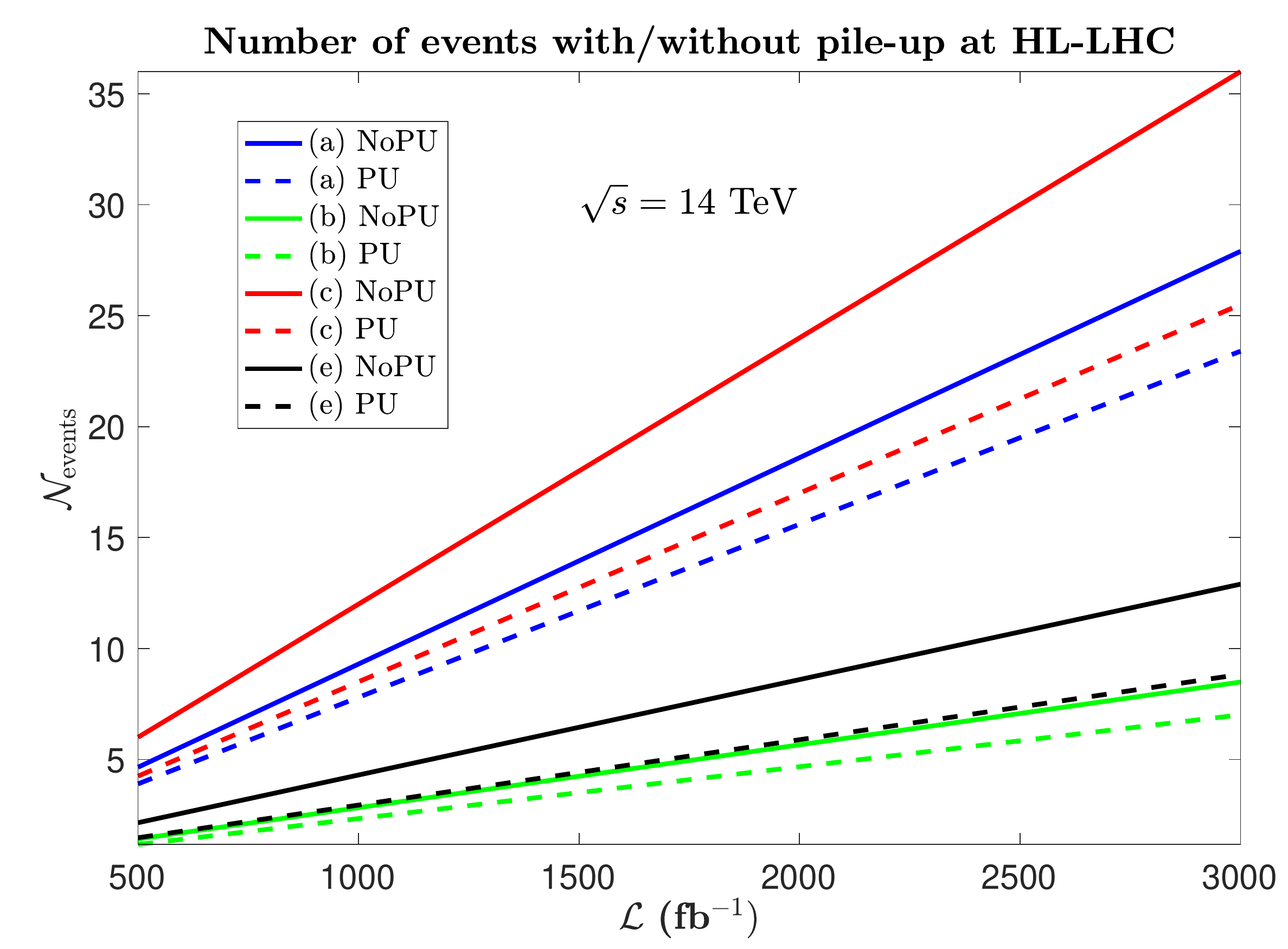}
\includegraphics[width=0.48\textwidth]{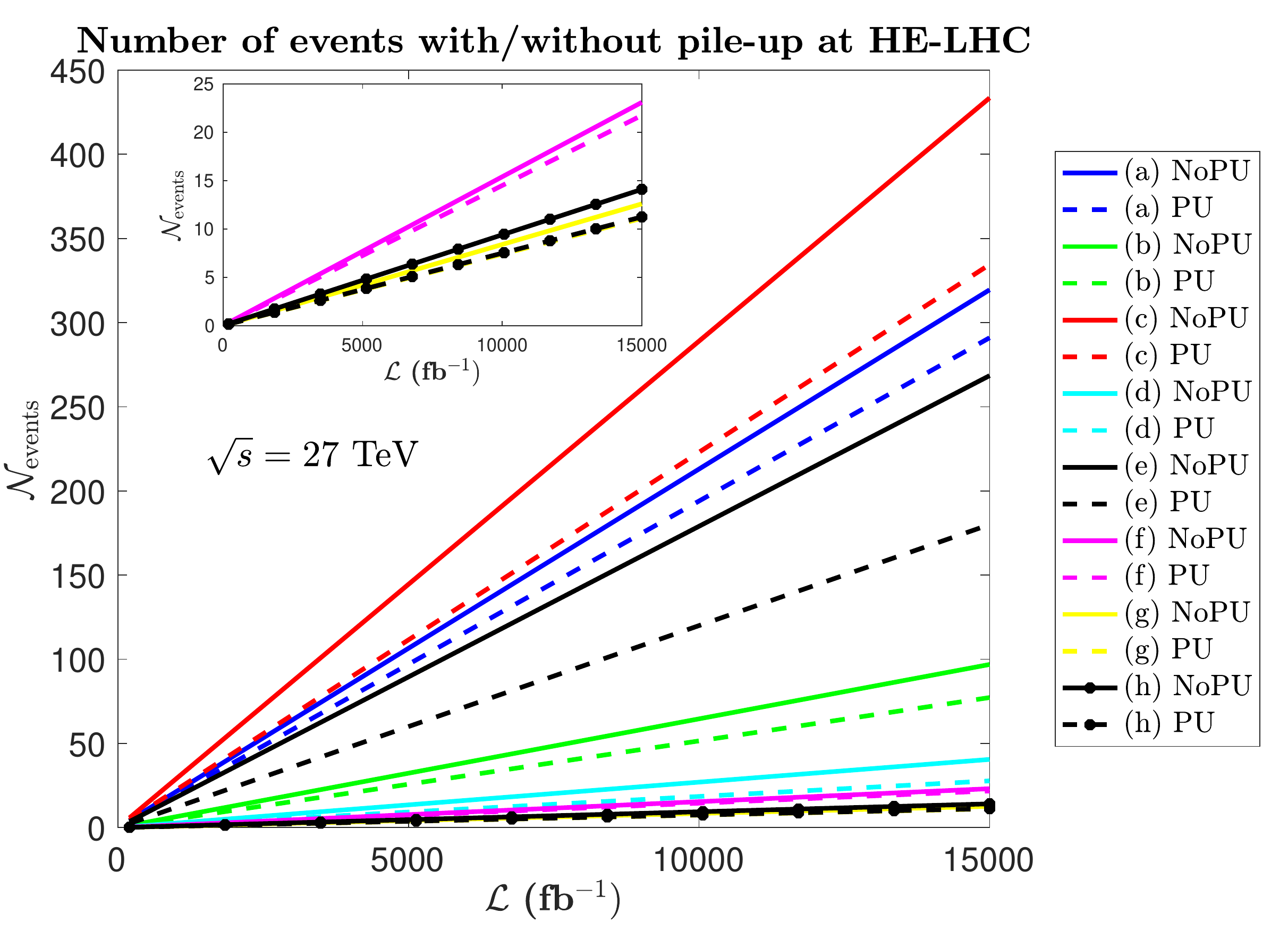}
\caption{Left panel: Estimated number of events for various integrated luminosities for benchmarks (a), (b), (c) and (e) in cases of no pile-up (solid lines) and pile-up (dashed lines) at HL-LHC. Right panel: same as the left panel but for HE-LHC for all the benchmarks of Table~\ref{tab1}.}
\label{fig4}
\end{figure}

Given the rate at which the HL-LHC is collecting data, points (a) and (c) may need a run time of $\sim 2$ years to be discovered while points (b) and (e) may take up to 5 to 6 years. This is shown in Fig.~\ref{fig5}. On the other hand, the run time should be greatly reduced at the HE-LHC which is expected to collect data at the rate of $820 \ifb$/year. Thus points (a) and (c) will require $\sim 4$ months of runtime while points (b) and (e) may take up to 5 months to a year. As for the other points, point (d) needs $\sim 3$ years, point (f) $\sim 5$ years and points (g) and (h) $\sim 8$ years.

\begin{figure}[H]
\centering
\includegraphics[width=0.5\textwidth]{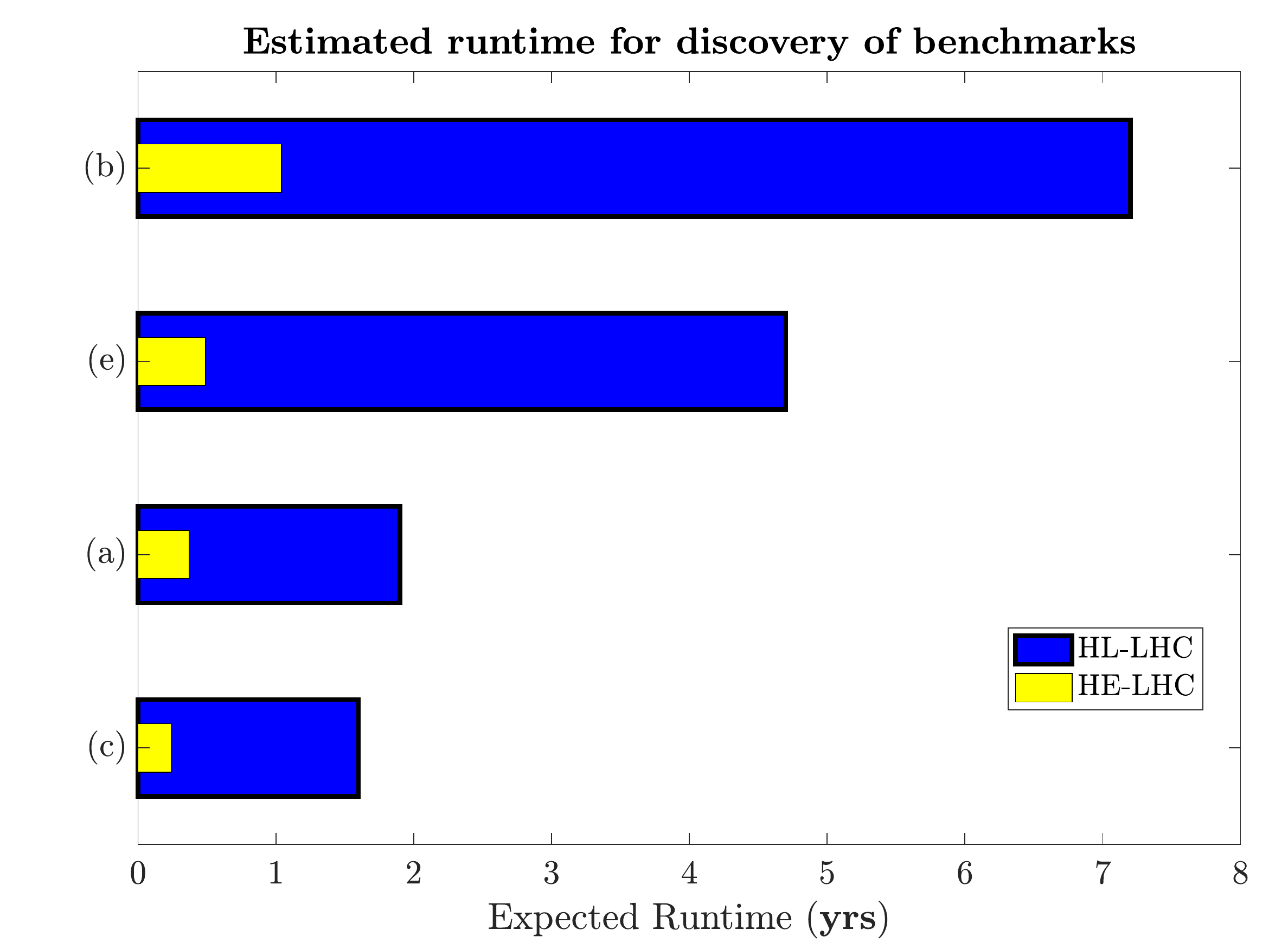}
\caption{Estimated runtime, in years, for the potential discovery of benchmark points (a), (b), (c), and  (d) that are within reach of both HL-LHC and HE-LHC. Blue bars represent HL-LHC and yellow bars are for HE-LHC. }
\label{fig5}
\end{figure}

Before concluding, we discuss the effect of the mass mixing coefficient $\epsilon$ and the kinetic mixing coefficient $\delta$ starting from the GUT scale. Thus in this analysis, we have set $\epsilon$ to zero (i.e. $M_2=0$) at the GUT scale and gave $\delta$ a non-zero value. From Eq.~(\ref{m2rge}), the RGE running of $M_2$ induces a tiny value for $\epsilon$ at the electroweak scale. Thus our analysis includes the effect of both mass and kinetic mixings. Now, one can reverse the situation and set $\delta$ to zero and give $\epsilon$ a tiny value at the GUT scale. The coefficient $\delta$ does not run and remains zero at the electroweak scale. We have checked that with some tuning of $\epsilon$ one can reproduce the same effect that $\delta$ has on the stau decay width. For example, if we consider point (a) of Table~\ref{tab1} and set $\delta=0$ and $\epsilon=6.2\times 10^{-7}$ at the GUT scale (so that $\epsilon=4.2\times 10^{-7}$ at the electroweak scale) we reproduce the same decay width and lifetime as given in Table~\ref{tab2} for the same point.

\section{Conclusions}\label{sec:conc}

In this analysis we presented an extension of the MSSM/SUGRA with an extra abelian gauge group $U(1)_X$. Under this extension, the MSSM/SUGRA is augmented by an additional $U(1)$ vector supermultiplet and a $U(1)$ chiral supermultiplet. The MSSM fields are not charged under $U(1)_X$ and the only communication between the MSSM and the hidden sector is through a gauge kinetic mixing coefficient, $\delta$ and a mass mixing parameter, $\epsilon$. As a result, the neutral gauge boson sector has an additional boson: a $Z'$ boson with couplings to the MSSM suppressed by $s_{\delta}$ and $\epsilon$ which can easily escape detection due to its very small production cross-section at colliders. The gaugino sector is extended as well and in particular the neutralino mass matrix becomes $6\times 6$ with two additional neutralinos. The lightest of the six neutralinos is the hidden sector, $\tilde\xi^0_1$ which is a dark matter candidate with very weak interaction with the visible sector. The NLSP is the stau which has a suppressed decay channel to $\tilde\xi^0_1$ making it a long-lived particle. The suppression is due to two sources: a small mixing coefficient, $\delta$, and a phase space suppression, i.e. a small mass gap between the LSP and the NLSP. Even though the dark matter candidate has very weak interactions with the bath, its relic abundance may still be produced via the freeze-out mechanism through conversion-driven processes. Since dark matter self-annihilation and stau-LSP annihilation are highly inefficient, the LSP relic density is mainly set by stau-stau annihilation to SM particles with the stau decay and inverse decay to the LSP responsible for maintaining chemical equilibrium. The strength of the latter process falls below the Hubble parameter at freeze-out for two of the considered benchmark points, (g) and (h), which makes co-scattering a leading process in converting $\tilde\xi^0_1$ to $\tilde\tau$ followed by stau self-annihilation which eventually depletes the relic abundance. Because of its very weak interactions, the LSP-proton scattering cross-section is negligible making such a dark matter candidate easily escape direct detection in scattering experiments. However, because of the suppressed decay width of the stau which is the NLSP, an opportunity to observe such a particle through its long-lived decay at the LHC exists and is of interest. The charged stau will leave a track in the ID before decaying into the hidden sector dark matter. Thus for this class of models, an observation of such a track will point to the existence of hidden sector dark matter.

In summary, we have given an analysis for the potential of HL-LHC and HE-LHC discovering hidden sector dark matter via long-lived stau through its pair production and its associated production with a tau sneutrino and its subsequent decay. The characteristic signature of a charged long-lived stau is a high $p_T$ track decaying to another charged track (resulting in a kinked track) with leptons having a large impact parameter. With proper cuts on select kinematic variables, all physical backgrounds can be rejected for both cases of no pile-up and pile-up. We show that half of the eight benchmark points considered can be discovered at the HL-LHC while all of those points are with in reach of the HE-LHC. It is also shown that a transition from HL-LHC to HE-LHC will reduce the runtime for discovery of points (a), (b), (c) and (e) by $\sim 80$ to $\sim 90\%$.

\vspace{1cm}

\textbf{Acknowledgments: }
We are grateful to Florian Staub for his help with \code{SARAH}, to Alexander Pukhov for many useful discussions regarding dark matter relic density calculations with \code{micrOMEGAs} and to Christian Ohm for insightful discussions on experimental signatures of LLPs. The analysis presented here was done using the resources of the high-performance  Cluster353 at the Advanced Scientific Computing Initiative (ASCI) and the Discovery Cluster at Northeastern University.  This research was supported in part by the NSF Grant PHY-1620575.

\end{document}